\documentclass[twocolumn,aps,prd,showpacs,nofootinbib,floatfix,eqsecnum]{revtex4}  
\usepackage{amsmath,latexsym,graphicx,multirow,longtable}     
\begin{document}  
\title{Binary black hole merger gravitational waves and recoil in the
large mass ratio limit}

\author{Pranesh A.\ Sundararajan${}^{1}$\footnote{PAS is currently a
Desk Strategist at Morgan Stanley and Co.\ Inc., 1585 Broadway, New
York, NY 10036.}}
\author{Gaurav Khanna${}^{2}$}
\author{Scott A.\ Hughes${}^{1}$}
  
\affiliation{${}^{1}$Department of Physics and MIT Kavli Institute,
MIT, 77 Massachusetts Ave., Cambridge, MA 02139 \\ ${}^{2}$Department
of Physics, University of Massachusetts, Dartmouth, MA 02747}
  
\date{\today}  
\begin{abstract}
Spectacular breakthroughs in numerical relativity now make it possible
to compute spacetime dynamics in almost complete generality, allowing
us to model the coalescence and merger of binary black holes with
essentially no approximations.  The primary limitation of these
calculations is now computational.  In particular, it is difficult to
model systems with large mass ratio and large spins, since one must
accurately resolve the multiple lengthscales which play a role in such
systems.  Perturbation theory can play an important role in extending
the reach of computational modeling for binary systems.  In this
paper, we present first results of a code which allows us to model the
gravitational waves generated by the inspiral, merger, and ringdown of
a binary system in which one member of the binary is much more massive
than the other.  This allows us to accurately calibrate binary
dynamics in the large mass ratio regime.  We focus in this analysis on
the recoil imparted to the merged remnant by these waves.  We closely
examine the ``antikick,'' an anti-phase cancellation of the recoil
arising from the plunge and ringdown waves, described in detail by
Schnittman et al.  We find that, for orbits aligned with the black
hole spin, the antikick grows as a function of spin.  The total recoil
is smallest for prograde coalescence into a rapidly rotating black
hole, and largest for retrograde coalescence.  Amusingly, this
completely reverses the predicted trend for kick versus spin from
analyses that only include inspiral information.
\end{abstract}  
\maketitle  
  
\section{Introduction and background}

\subsection{Modeling binary systems in general relativity}

After roughly three decades of effort, numerical relativity can now
model nearly arbitrary binary black hole configurations.  Following
Pretorius' pioneering ``breakthrough'' calculation {\cite{frans1}},
and then the successes of the Brownsville and Goddard groups using
techniques that required only modest modifications to the methods they
used before the breakthrough {\cite{brownsville,goddard}}, the past
few years have seen an explosion of activity.  Recent work has studied
the impact of the many physical parameters which describe binaries,
such as mass ratio {\cite{massratio1,massratio2}}, spin and spin
alignment {\cite{spin1, spin2, spin3, spin4}}, and eccentricity
{\cite{ecc1,ecc2}}.  As numerical models have improved, analytic tools
for modeling binary systems {\cite{blanchet06}} and connecting
numerics and analytics have likewise matured.  In particular, the {\it
effective one-body} (EOB) {\cite{bd1, bd2, djs, d01}} approach to
binary dynamics, which maps the dynamics of a binary to that of a
point particle moving in an ``effective'' spacetime corresponding to a
deformed black hole, has been found to outstandingly describe the
outcome of numerical relativity calculations after some adjustable
parameters in the EOB framework are calibrated to numerical
calculations {\cite{eob_num1, eob_num2, eob_num3, eob_num4}}.  Our
understanding of the two-body problem in general relativity has never
been better.

These efforts are largely motivated by the need for accurate models of
coalescing black holes to detect and measure merger signals in the
data of gravitational-wave (GW) detectors.  Black holes with masses of
roughly $10^6 - 10^9 \,M_\odot$ indisputably reside at the cores of
essentially every galaxy with a central bulge
{\cite{kg01,ferrarese02}}.  In the hierarchical growth of structure,
these black holes will form binaries as their host galaxies merge and
grow {\cite{bbr1980}}; estimates of how often such binaries form
indicate that the proposed space-based detector LISA
{\cite{lisa_nasa,lisa_esa}} should be able to measure at least several
and perhaps several hundred coalescences over a multiyear mission
lifetime {\cite{bbh_rate_lisa}}.  There is already a catalog of
candidate binaries in this mass range, such as active galaxies with
double cores \cite{komossa2003, maness2004, rodrigues2006}, systems
with doubly-peaked emission lines \cite{zhou2004, gerke2007}, and
systems that appear to be periodic or semi-periodic, such as the
blazar OJ287 \cite{valtonen2007}.  The last year or so of the binary's
life will generate GWs at frequencies to which LISA is sensitive;
measuring those waves will make it possible to precisely map the
distribution of cosmic black hole masses and spins, opening a new
observational window onto the high-redshift growth of cosmic
structure.

Less massive black hole binaries (several to several hundred
$M_\odot$) will be targets for the ground-based GW detector network,
currently including LIGO {\cite{ligo}}, Virgo {\cite{virgo}}, and GEO
{\cite{geo}}, and hopefully including the proposed detectors LCGT
{\cite{lcgt}}, AIGO {\cite{aigo}}, and the ``Einstein Telescope''
{\cite{et}} in the future.  Formation scenarios and event rate
estimates in this band are much less certain, since the demographics
of the relevant black holes and scenarios for them to form binaries
are not as well understood as in the supermassive range.  However,
scenarios involving dynamic binary formation in dense clusters suggest
that this network can plausibly expect an interesting event rate
{\cite{bbh_rate_ligo1, bbh_rate_ligo2, bbh_rate_ligo3, bbh_rate_ligo4,
bbh_rate_ligo5}}, strongly motivating the construction of binary
merger models for these detectors.

Finally, moving back to the LISA band, binaries in which one member is
much less massive than the other are expected to be an important
source.  Such extreme mass ratio binaries are created when a stellar
mass secondary ($\sim 1 - 100\,M_\odot$) is scattered through
multibody interactions onto a highly relativistic orbit of a roughly
$10^6\,M_\odot$ black hole in the center of a galaxy.  Though rare on
a galaxy-by-galaxy basis, enough galaxies will be in the range of LISA
that the number measurable is expected to be several dozen to several
hundreds {\cite{emri_rate}}.  The waves from these binaries largely
probe the quiescent spacetimes of their larger (presumably Kerr) black
hole, making possible precision tests of the strong-field nature of
black hole spacetimes {\cite{emri_science}}.

In short, astrophysical binary black holes will come in a wide range
of mass ratios.  Computational models must be able to handle systems
with mass ratios ranging from near unity, to millions to one.  Each
mass $m$ sets a lengthscale $Gm/c^2$ which the code must be able to
resolve.  Large mass ratios require codes that can handle a large
dynamic range of physically important lengthscales.

Perturbation theory is an excellent tool for modeling binaries with
very large mass ratios.  In this limit, the binary's spacetime is
nearly that of its largest member, with the smaller member acting to
distort the metric from the (presumably) exact Kerr solution of that
``background.''  It is expected that tools based on perturbation
theory will be crucial for modeling extreme mass ratio systems
described above (mass ratios of $10^4:1$ or larger).  Even for less
extreme systems, perturbative approaches are likely to contribute
important wisdom, working in concert with tools such as numerical
relativity and the effective one-body approach.

The foundational examples of such an analysis are the papers of Nagar,
Damour, and Tartaglia {\cite{ndt07}} and of Damour and Nagar
{\cite{dn07}}.  In that work, the EOB framework is used to construct
the quasi-circular late inspiral and plunge of a small body into a
non-rotating black hole.  Regge-Wheeler-Zerilli methods
{\cite{rw57,z70}} are then used to compute the GWs that arise from a
small body that follows that trajectory into the larger black hole.
Those authors use this large mass ratio system as a ``clean
laboratory'' for investigating binary dynamics, and advocate using
these techniques as a tool for probing delicate issues such as the
form of the waves which arise from the plunge, and the matching of the
final plunge waves to the late ringdown dynamics of the system's final
black hole.

Our goal here is to develop a similar toolkit based on perturbation
theory applied to spinning black holes.  We have developed two
perturbation theory codes which we use to model different aspects of
binary coalescence.  Both codes solve the Teukolsky equation
{\cite{teuk}}, computing perturbations to the curvature of a Kerr
black hole.  One code works in the frequency domain {\cite{h2000,
dh2006}}, which works well for computing the averaged flux of
quantities such as energy and angular momentum carried by GWs.  The
other code works in the time domain {\cite{skh07, skhd08}}, which is
excellent for calculating the aperiodic GW signature of an evolving
source.  As originally proposed in Ref.\ {\cite{hdff05}}, we have
developed a hybrid approach which uses the best features of both the
time- and frequency-domain codes to model the full coalescence
process.  (Although our ultimate goal is to develop a set of tools
similar to those developed by Damour, Nagar, and Tartaglia, we note
that our techniques are the moment largely numerical, as opposed to
the mixture of numerical and analytic techniques developed in Refs.\
{\cite{ndt07,dn07}}.  It would be worthwhile to connect the work we
present here to the body of EOB work, but have not yet begun doing so
in earnest.)

As we were completing this paper, a perturbation-theory-based analysis
of binary merger was presented by Lousto et al.\ {\cite{lnzc}}.  Their
analysis does not use the Teukolsky equation, but is otherwise very
similar in style and results to what we do here.  In particular, they
note as we do here that the perturbation equations terminate the
merger waveform in a set of ringdown waves in a very natural way,
thanks to the manner in which the equation's source redshifts away as
the infalling body approaches the large black hole's event horizon.
This behavior was also pointed out and exploited by Mino and Brink
{\cite{mb08}} in their (largely analytic) perturbative analysis of
recoil from waves from the late plunge.  We expand on this point in
more detail at appropriate points later in the paper.

As our use of the Teukolsky equation requires, we assume that a binary
can be well-described by a small body moving in the spacetime of a
(much larger) Kerr black hole.  We first build the worldline that the
smaller body follows as it slowly inspirals and then plunges into the
black hole.  We assume that, early in the coalescence, the small body
moves on a geodesic of the background Kerr spacetime.  Using the
frequency-domain perturbation theory code to compute their rates of
change, we allow the energy $E$, angular momentum $L_z$, and Carter
constant $Q$ of this configuration to evolve.  (In fact, we confine
ourselves to equatorial orbits in this analysis, so $Q = 0$ throughout
the binary's evolution.)  This drives the smaller body in an adiabatic
inspiral through a sequence of orbits, until we approach the last
stable orbit of the large black hole\footnote{At present, we do not
include the conservative impact of self forces.  These forces {\it
are} included in the EOB-framework analyses of Damour, Nagar, and
Tartaglia.}.

We then make a transition to a plunging orbit, using the prescription
of Sundararajan {\cite{s2008}} which in turn generalized earlier work
by Ori and Thorne\footnote{A similar approach to the transition from
inspiral plunge, but valid for arbitrary mass ratios and presented
using the EOB framework, was developed by Buonanno and Damour
{\cite{bd2}}, and appeared in press before Ref.\ {\cite{ot2000}}; we
thank T.\ Damour for clarifying this chronology to us.  This approach
is used in Refs.\ {\cite{ndt07,dn07}} to compute the transition from
the slow, adiabatic inspiral to plunge.} {\cite{ot2000}}.  By properly
connecting the adiabatic inspiral to a plunge, we make a full
worldline describing the small body's coalescence with the larger
black hole.  This worldline gives us the source for our time-domain
perturbation theory code, from which we compute the GWs generated by
the system as the small body evolves from the (initially near
geodesic) inspiral through the plunge and merger.  The waves which we
compute in this way have qualitatively the same ``inspiral, merger,
ringdown'' structure seen in numerical relativity simulations, though
much work remains to quantify the degree of overlap.

As an illustration of the utility of our perturbative toolkit, we
focus in this paper on the problem of GW recoil.  Studies of GW recoil
have been particularly active in recent years; we review this problem
and its literature in the next subsection.

\subsection{Gravitational-wave recoil}

The asymmetric emission of GWs from a source carries linear momentum.
The system then recoils to enforce global conservation of momentum.
Early work demonstrated the principle of this phenomenon
\cite{br1961,peres1962}; Bekenstein \cite{b1973} appears to have been
the first to appreciate the important role it could play in
astrophysical problems.  Much recent work has focused on the recoil
imparted to the merged remnant of binary black hole coalescence.

The first estimates of binary black hole kick were made by Fitchett
\cite{fitchett}.  He treated the gravitational interaction as
Newtonian and included the lowest order mass and current multipoles
needed for GW emission to compute the recoil velocity.  This early
calculation predicted that recoil velocities could approach thousands
of km/s, which is greater than the escape velocity for many galaxies.
Because of his restriction to low-order radiation formulas, and his
use of Newtonian gravity to describe binary dynamics, it was clearly
imperative that Fitchett's calculations be revisited; a prescient
analysis by Redmount and Rees {\cite{redmountrees}} particularly
argued for the need to account for the effect of black hole spins in
the coalescence.

Over the past several years, quite a few calculations have
substantially improved our ability to model the recoil in general
relativity.  The various approaches can be grouped as follows:

\begin{itemize}

\item {\it Black hole perturbation theory}.  As discussed extensively
above, black hole perturbation theory is a good tool for describing
binaries involving a massive central black hole (of mass $M$) and a
much less massive companion (of mass $\mu$).  Shortly after Fitchett's
pioneering binary calculation, Fitchett and Detweiler examined whether
strong-field gravity changed the conclusions using perturbation theory
{\cite{fd84}}.  Twenty years later, Favata, Hughes, and Holz
\cite{FHH} argued that, properly extrapolated, reasonable results can
be obtained for quantities such as the integrated black hole kick up
to a mass ratio $\mu/M \sim \mathcal O(0.1)$.  Unfortunately, the
Favata et al.\ analysis has a rather large final error since the
frequency-domain tools they use do not work well at modeling the GWs
arising from the final plunge of the smaller body into the large black
hole.  One of our goals in this analysis is to revisit that
calculation and reduce those substantial error bars.

Another application of perturbation theory is the ``close-limit
approximation,'' {\cite{closelimit}} which describes the last stages
of a merging binary as the dynamics of a distorted single black hole.
Sopuerta, Yunes, and Laguna \cite{syl06} applied the close-limit
approximation to describe the final waves from unequal mass binaries,
obtaining results that compare very well with those that have since
been computed within ``full'' numerical relativity.

Finally, Mino and Brink {\cite{mb08}} used perturbative techniques to
model the waves from the plunge, quantifying the manner in which the
geometry of the final infall impacts the kick imparted to the binary.
As already mentioned, their analysis also took advantage of the manner
in which the source redshifts away as the infalling body approaches
the larger black hole's event horizon.

\item {\it Post-Newtonian (PN) theory}.  PN theory describes the
spacetime and the motion of bodies in the spacetime as an expansion in
the Newtonian gravitational potential $Gm/rc^2$ (where $m$ is a
characteristic system mass, and $r$ a characteristic black hole
separation).  Blanchet, Qusailah, and Will \cite{bqw05} used an
approach based on this expansion to substantially improve estimates
from the recoil from the final plunge and merger; though consistent
with the results from {\cite{FHH}}, they were able to reduce the error
bars by a substantial factor.  More recently, Le Tiec, Blanchet, and
Will {\cite{ltbw10}} combined a post-Newtonian inspiral with a
close-limit computation of the merger and ringdown to compute the
recoil for the coalescence of non-spinning black holes.  This analysis
is quite similar in spirit to the one we present here, though it does
not use perturbation theory throughout.

\item {\it Numerical relativity}.  Not long after it first became
possible to model the coalescence of two black holes in numerical
relativity, this became the technique of choice for computing black
hole recoil.  No other technique is well-suited to computing wave
emission and spacetime dynamics for very asymmetric, strong-field
configurations which are likely to produce strong GW recoils.
Numerical relativity was needed to discover the so-called
``superkick'' configuration: an alignment of spin and orbital angular
momentum which results in a kick of several thousand kilometers per
second {\cite{superkick0,superkick1,superkick2}}.  In most
configurations, the kick tends to be substantially smaller, peaking at
a few hundred kilometers per second \cite{NR1,NR2,NR3}.
 
\item {\it Effective one-body}.  As already described, EOB describes a
binary as a test body orbiting in the spacetime of a ``deformed''
black hole, with the deformation controlled by factors such as the
mass ratio of the binary.  Damour and Gopakumar {\cite{dg06}} first
examined the issue how to compute recoil within the EOB framework,
analytically identifying the major contributions to the recoil that
accumulates over a coalescence, including the importance of the final
merger and recoil waves in providing an ``antikick'' contribution.  By
calibrating some parameters of the EOB framework with results from
numerical relativity, EOB has had great success generating waveforms
and recoil velocities that match well with those from numerical
relativity \cite{EOB_orig,EOB}.

\end{itemize}
With the exception of the ``superkick'' configuration, all of these
techniques predict recoils that peak at roughly a few hundred
kilometers per second (depending on mass ratio, spins, and spin orbit
orientation; see {\cite{statistics}} for detailed discussion and
statistical analysis).  This is substantially lower than the peak
predicted by Fitchett's original calculation; his overestimate can be
ascribed to neglect of important curved spacetime radiation emission
and propagation effects.

In addition to their potential astrophysical applications, recoil
computations serve another important purpose: They are a common point
of comparison for these four approaches to strong field gravity.  The
recoil velocity from a merging binary is calculated by integrating the
emitted radiation over some number of orbits.  Any significant
systematic error in the approach used used will tend to magnify the
error in the estimated recoil velocity.  Thus, the evaluated recoils
for a range of BH spins and mass ratios serve as a good platform for
comparing various approaches to strong field binary models.

\subsection{This paper}

Our goal is to revisit and improve the estimate of black hole recoil
via black hole perturbation theory that was originally developed in
Ref.\ {\cite{FHH}}.  That analysis predicts upper and lower bounds
which are rather widely separated.  This is because the analysis of
{\cite{FHH}} could not accurately model wave emission from the final
plunge and merger.  Using the time-domain perturbation theory code
developed and presented in Refs.\ {\cite{skh07,skhd08}}, we can now
compute the contribution of those waves.  As we describe in more
detail in Sec.\ {\ref{sec:results}}, doing so completely reverses the
conclusions of Ref.\ {\cite{FHH}} regarding how the kick behaves as a
function of spin.  In particular, including the plunge and merger is
crucial to correctly computing the ``antikick,'' the out-of-phase
contribution to the recoil that arises from the merger's final GWs.
This contribution to a binary's total recoil was first identified and
characterized by Schnittman et al.\ {\cite{antikick}}.  We find that
the inability to include this contribution in Ref.\ {\cite{FHH}} is
largely responsible for the large error bars in that analysis.

We begin by reviewing in Sec.\ {\ref{sec:trajectory}} how we construct
the worldline which the smaller member of our binary follows as it
spirals into the larger black hole.  As briefly described above, we
break this trajectory into a slowly evolving ``inspiral'' (Sec.\
{\ref{sec:inspiral}}) followed by a transitional regime (Sec.\
{\ref{sec:transition}}) that takes the binary into a final plunge and
merger (Sec.\ \ref{sec:plunge}).  This review is left general, so that
in principle one could describe this dynamics for generic orbital
geometry.  We specialize in our analysis here to the simplest circular
and equatorial orbits (Sec.\ {\ref{sec:circeq}}).

We next briefly review how we compute gravitational radiation from a
body moving on this trajectory.  As mentioned above, our approach is
based on finding solutions to the Teukolsky equation {\cite{t73}} for
Kerr black hole perturbations.  We review this equation's general
properties in Sec.\ {\ref{sec:radiation}}, and then discuss the
principles behind solving it in the frequency domain (Sec.\
{\ref{sec:freqdomain}}) and in the time domain (Sec.\
{\ref{sec:timedomain}}).  Section {\ref{sec:recoil}} summarizes how
one computes the radiation's linear momentum and the recoil of a
merged system.

Section {\ref{sec:results}} presents the results of our analysis.  We
begin in Sec.\ {\ref{sec:massratio}} with general considerations on
how our results scale with mass ratio.  Because we work strictly
within the context of linearized perturbation theory, all of our
results can be easily scaled to different mass ratios, provided that
the scaling does not change the system so much that the validity of
perturbation theory breaks down.  Reference {\cite{FHH}} argued that a
modified scaling would allow us to estimate with reasonable accuracy
quantities related to the recoil even out of the perturbative regime.
Although those arguments are valid during the adiabatic inspiral, they
break down when the members of the binary merge.

In Sec.\ {\ref{sec:waves}}, we then discuss in some detail the
gravitational waveform we find for binary coalescence in the large
mass ratio limit.  We examine the different multipolar contributions
to the last several dozen cycles of inspiral, followed by the plunge
and merger.  These examples illustrate the manner in which the
coalescence waves very naturally evolve into a ``ringdown'' form.  As
discussed in some detail in Sec.\ {\ref{sec:timedomain}}, this
behavior arises by virtue of how the Teukolsky equation's source term
goes to zero, so that its solutions transition to their homogeneous
form, as the infalling body approaches the large black hole's event
horizon.  Mino and Brink {\cite{mb08}} first appear to have exploited
this behavior, which was also seen in recent work by Lousto, Nakano,
Zlochower, and Campanelli {\cite{lnzc}}.  This demonstrates the power
of perturbative methods at modeling physically important aspects of
the merger waves.

Section {\ref{sec:spin}} examines the recoil that arises from these
waves, focusing on how it depends (for the circular, equatorial case
that we study) on the spin of the larger black hole.  This analysis
demonstrates very clearly the impact of the ``antikick'' first
reported by Schnittman et al.\ {\cite{antikick}}.  With the antikick
taken into account, the smallest recoils come from the largest spins
when the merger is a prograde sense; the largest spins come from
retrograde mergers with large spins.  The waves which give the system
its antikick come from those produced by the final plunge and merger,
demonstrating very clearly the substantial impact these waves have on
the system.  We conclude this section by briefly discussing the
convergence of our recoil results as a function of black hole spin.
Interestingly, we find that the number of modes we must include in
order for our results to converge is a strong function of the black
hole's spin --- rapid spin, prograde cases need more modes than do
slow spin cases, which in turn need more modes than rapid spin,
retrograde cases.  We conclude the paper by discussing in Sec.\
{\ref{sec:conclude}} how these tools may be used to expand the reach
of two-body modeling in general relativity, and our future plans.

Throughout our analysis we generally use units in which $G = c = 1$.
We sometimes use $c = 3\times10^5\,{\rm km/sec}$ in order to present
kicks in ``physical'' units.
 
\section{Building the inspiral and plunge trajectory}
\label{sec:trajectory}

Roughly speaking, our coalescence model has two ingredients.  First,
we compute the worldline that the small body follows as it spirals
from large radius through plunge into the black hole.  We then use
that worldline to build the source for the Teukolsky equation and
compute the GWs that are generated as the smaller body follows the
worldline into the black hole.  Though for simplicity we describe
these ingredients as though they stand in isolation, they are in fact
strongly coupled.  We describe here how we compute the inspiral and
plunge trajectory, deferring discussion of how we compute radiation
from this trajectory to Sec.\ {\ref{sec:radiation}}.  Throughout, we
indicate how these steps are coupled to one another.

The trajectory which the small body follows can be broken into three
pieces: An early time {\it inspiral}, in which the smaller member of
the binary is approximated as evolving through a sequence of bound
orbits of the larger black hole; a late time {\it plunge}, in which
the small body falls into the larger black hole; and an intermediate
{\it transition} which smoothly connects these two regimes.  We now
briefly review how we model these different pieces.

In all of these regimes, we treat the zeroth order motion of the small
body as a geodesic of the Kerr spacetime.  These geodesics must be
augmented by the conservative action of a self force if one's goal is
to make a model that faithfully reproduces the phase of binary black
hole GWs.  For our present goal of estimating the GW recoil, we expect
that the error due to neglecting this force is not important.  Kerr
black hole geodesics \cite{bpt72} are described by the following
equations for the motion in Boyer-Lindquist coordinates $r$, $\theta$,
$\phi$, and $t$:
\begin{eqnarray}
\label{eq:geod1}
 \Sigma \frac{dr}{d\tau} &=& \pm \sqrt{R} \; , \\
\label{eq:geod2} 
\Sigma \frac{d\theta}{d\tau} &=& \pm \sqrt{V_\theta} \; , \\
\label{eq:geod3} 
\Sigma \frac{d\phi}{d\tau} &=& V_\phi \; , \\
\label{eq:geod4} 
\Sigma \frac{dt}{d\tau} &=& V_t \;.
\end{eqnarray}
The potentials appearing here are
\begin{eqnarray}
R & = &   \left[E \left(a^2+r^2\right)-a L_z\right]^2
\nonumber\\
& &\qquad -\Delta\left[(L_z-a E)^2 + \mu^2 r^2 + Q\right] \;,
\label{eq:Vr}\\
V_\theta & = & Q - \cos^2\theta\left[a^2\left(\mu^2-E^2\right) +
\csc^2\theta L_z^2\right] \;,
\label{eq:Vth}\\
V_\phi & = & \csc^2\theta L_z - a E 
 + \frac{a}{\Delta}\left[E\left(r^2 + a^2\right)-L_z a\right] \;,
\label{eq:Vph}\\
V_t & = & a\left(L_z - a E \sin^2\theta\right)
\nonumber\\
& &\qquad +\frac{r^2 + a^2}{\Delta}\left[E\left(r^2 + a^2\right) - L_z
a\right] \;.
\label{eq:Vt}
\end{eqnarray}
The quantity $M$ is the large black hole's mass, $a$ is that hole's
Kerr spin parameter, and $\mu$ is the mass of the smaller body which
perturbs the black hole spacetime.  The functions $\Sigma = r^2 + a^2
\cos^2\theta$ and $\Delta = r^2 - 2 M r + a^2$.  In the absence of
radiation emission, the energy $E$, axial angular momentum $L_z$, and
Carter constant $Q$ are constants of the motion; up to initial
conditions, choosing these three constants defines a geodesic.

Equations (\ref{eq:geod1}) -- (\ref{eq:geod4}) are the starting point
for building the smaller body's inspiral and plunge worldline.  We now
describe in some detail how we use them for this computation.

\subsection{The inspiral}
\label{sec:inspiral}

We approximate the inspiral as a slowly evolving sequence of bound
Kerr geodesics (neglecting for now conservative aspects of the self
interaction).  Momentarily ignore the impact of radiation emission.
In this limit, the orbits are determined by selecting $E$, $L_z$, and
$Q$ plus initial conditions, and are completely characterized by three
orbital frequencies describing their periodic motions in the $r$,
$\theta$, and $\phi$ coordinates {\cite{schmidt}}.  This periodic
nature means that functions built from the orbital motion can be
usefully represented by a discrete Fourier expansion.

To build our inspiral, we assume that radiation acts slowly enough
that, to a good approximation, we can treat the small body's worldline
as a Kerr geodesic at each moment.  We then use the frequency-domain
Teukolsky solver described in Sec.\ {\ref{sec:radiation}} to compute
the rates at which $E$, $L_z$, and $Q$ evolve due to GW backreaction.
From these rates of change, we build the time-varying parameters
$E(t)$, $L_z(t)$, and $Q(t)$ which describes the sequence of orbits
the small body passes through on its inspiral.  More detailed
discussion of this procedure is given in Refs.\ {\cite{hdff05}} and
{\cite{h2001}}.

\subsection{The last stable orbit and the transition to plunge}
\label{sec:transition}

Our assumptions, and hence our procedure for computing the inspiral,
break down as the small body approaches the {\it last stable orbit},
or LSO.  This is worth describing in some detail.  For bound orbits,
the function $R(r)$ defined in Eq.\ (\ref{eq:Vr}) generally has four
real roots.  Denote these roots $r_1 > r_2 > r_3 > r_4$.  The root
$r_4$ is generally inside the event horizon\footnote{In fact, $r_4 =
0$ for equatorial orbits ($Q = 0$) and for orbits of Schwarzschild
black holes ($a = 0$).}, and is not interesting for our discussion.
The roots $r_1$, $r_2$, and $r_3$ on the other hand, are quite
important.  When these roots are distinct, the geodesic describes an
eccentric orbit that oscillates between $r_1$ (apoapsis) and $r_2$
(periapsis).  When $r_1 = r_2 > r_3$, the geodesic describes a
circular orbit at $r = r_1$.  (In this case, we also have $dR/dr = 0$
at $r = r_1$.)  When $r_2 = r_3$, the orbit is {\it marginally
stable}.  (The triple root $r_1 = r_2 = r_3$ denotes a marginally
stable circular orbit.)  Once we reach this point, the small body will
rapidly plunge into the black hole.  This condition defines the LSO.

As inspiral proceeds, the roots $r_2$ and $r_3$ approach one another,
indicating that GW backreaction is carrying the small body toward the
LSO.  We model the transition from slowly evolving geodesics through
the LSO to plunge by expanding the equations of motion around their
behavior at the LSO, as described in Ref.\ {\cite{s2008}} (which
generalizes Ref.\ \cite{ot2000}).  More specifically, we take the
constants in the transition to be given by
\begin{equation}
E(t) \simeq E_{\rm LSO} + (t - t_{\rm LSO})\dot E_{\rm LSO}\;,
\end{equation}
and similarly for $L_z$ and $Q$.  Here, $E_{\rm LSO}$ and $\dot E_{\rm
LSO}$ are the energy and its rate of change at the LSO (the latter
calculated using our frequency-domain Teukolsky equation solver), and
$t_{\rm LSO}$ is the time at which the LSO is reached.  We integrate
the geodesic equations using this form from a time $t_{\rm start} <
t_{\rm LSO}$ until a time $t_{\rm end} > t_{\rm LSO}$.  Reference
{\cite{s2008}} describes how we choose $t_{\rm start}$ and $t_{\rm
end}$ as a function of parameters such as the black hole spin $a$ and
binary mass ratio.  For our purposes, it is enough to note that,
provided they are chosen within a well-defined range, our results are
robust to that choice --- varying $t_{\rm start}$ and $t_{\rm end}$
does not significantly change the recoil.  A more careful
investigation may clarify the optimal way to define these transition
parameters.

\subsection{The plunge}
\label{sec:plunge}

For $t > t_{\rm LSO}$, the geodesics described by $E(t)$, $L_z(t)$,
and $Q(t)$ correspond to plunging geodesics, i.e., trajectories which
fall into the large black hole.  As described in Ref.\ {\cite{s2008}},
the transition matches onto a plunging trajectory most simply by just
holding these parameters constant for $t \ge t_{\rm end}$.  This is
justified by the fact that radiation reaction does not have a strong
impact in the final plunge {\cite{bd_mg,ndt07,dn07}}: careful
analysis indicates that an orbit's energy and angular momentum remain
nearly constant during the final plunge into the black hole.

As the small body approaches the black hole, its motion as viewed by
distant observers appears to ``freeze'' onto the generators of the
event horizon\footnote{This behavior led many researchers to call
these solutions ``frozen stars'' in the early literature.}.  When this
happens, the source term of the Teukolsky equation redshifts to zero
[cf.\ Eq.\ (2.46) of Ref.\ {\cite{skh07}}].  Since the homogeneous
Teukolsky equation's solutions are the quasi-normal modes of the
binary's large black hole, this means that the final cycles of
radiation from our coalescing system are very naturally given by the
system's ringdown modes.

\subsection{Specialization to circular equatorial orbits}
\label{sec:circeq}

Until now, we have kept the discussion of these inspiral and plunge
trajectories general in order to emphasize that our approach can be
applied to totally generic coalescences.  For this first analysis, we
now focus on the simplest interesting case, circular orbits confined
to the equatorial plane of the larger black hole.  In this limit,
geodesic orbits are totally characterized by the orbit's radius; Ref.\
{\cite{bpt72}} gives an outstanding summary of their properties.

\begin{figure*}[hbt]
\includegraphics[width = 8cm]{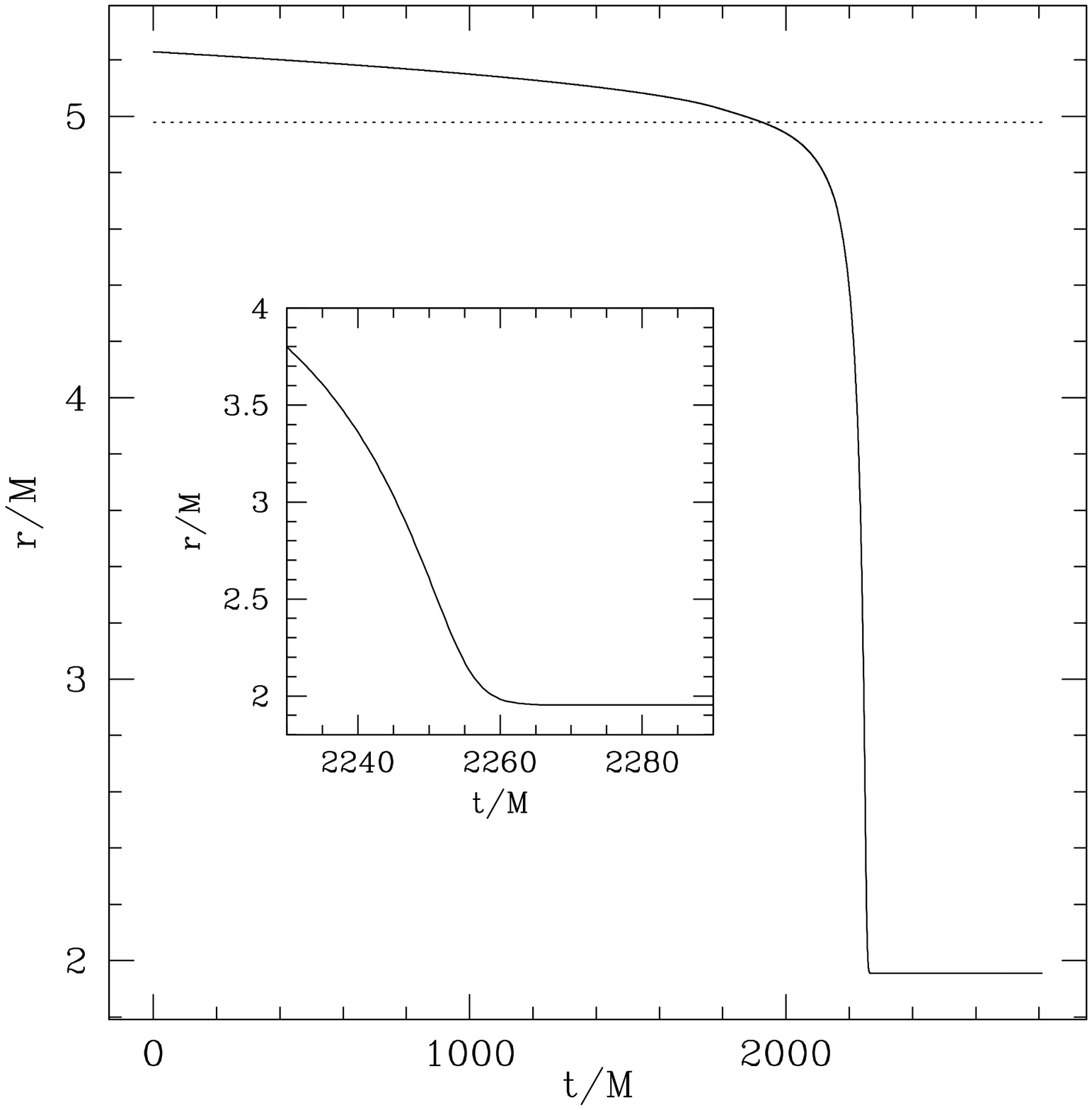}
\includegraphics[width = 8cm]{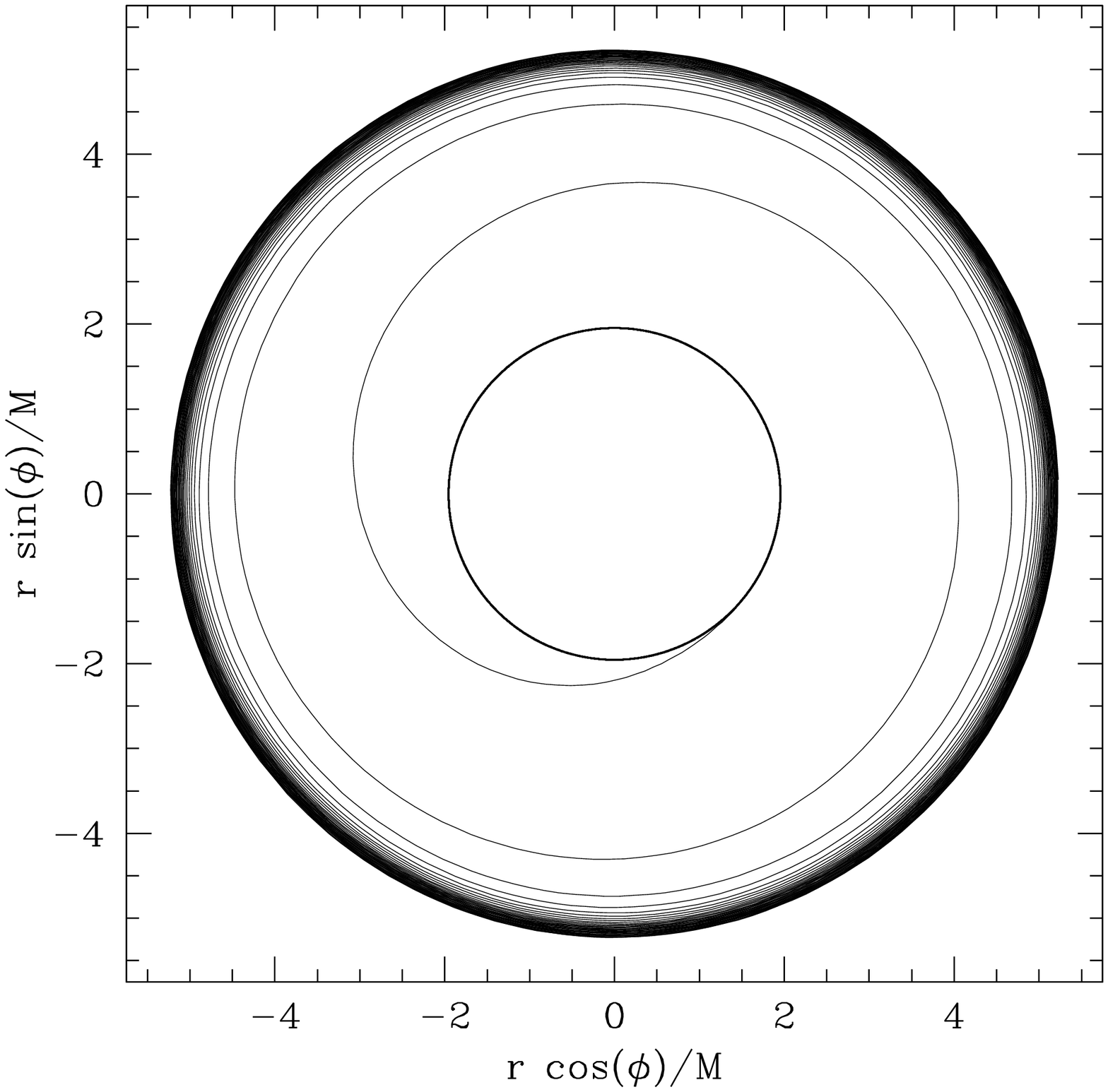}
\caption{Example quasi-circular inspiral and plunge trajectory.  For
this calculation, the larger black hole's spin was set to $a = 0.3M$,
and the binary's mass ratio is $\mu/M = 10^{-4}$.  Left panel shows
$r(t)$, the trajectory's radius as a function of time; right panel
shows the same trajectory as viewed in the equatorial plane of the
larger black hole.  On the left, the dotted line at $r = 4.98M$ labels
the radius of the prograde last stable orbit.  Our trajectory, which
starts at $r = 5.23M$, executes about 25 orbits before crossing this
point.  The inset there zooms in on the region $t \simeq 2260M$,
showing the smaller body's trajectory as it approaches the event
horizon at $r = 1.955M$.  On the right, the heavy circle at $r =
1.955M$ is the hole's event horizon; notice how the particle quickly
``locks'' onto the horizon after the plunge which follows its slow
inspiral.}
\label{fig:traj}
\end{figure*}

The energy, angular momentum, and Carter constant of circular
equatorial orbits are given by
\begin{eqnarray}
E &=& \frac{1 - 2M/r \pm a M^{1/2}/r^{3/2}}{\sqrt{1 - 3M/r \pm 2 a
M^{1/2}/r^{3/2}}}\;,
\label{eq:Eeqcirc}\\
L_z &=& \pm \frac{\sqrt{r M}\left(1 \mp 2 a M^{1/2}/r^{3/2} +
a^2/r^2\right)}{\sqrt{1 - 3M/r \pm 2 a M^{1/2}/r^{3/2}}}\;,
\label{eq:L_zeqcirc}\\
Q &=& 0\;.
\label{eq:Qeqcirc}
\end{eqnarray}
Upper sign refers to prograde coalescences (orbital angular momentum
parallel to black hole spin), lower sign to retrograde (antiparallel).
These orbits are characterized by a single frequency associated with
the azimuthal motion,
\begin{equation}
\Omega = \Omega_\phi = \pm\frac{M^{1/2}}{r^{3/2} \pm a M^{1/2}}\;.
\label{eq:omega_phi}
\end{equation}
The last stable orbit is located at
\begin{eqnarray}
r_{\rm lso}/M &=& 3 + Z_2 \mp \sqrt{(3 - Z_1)(3 + Z_1 + 2
Z_2)}\;,
\nonumber\\
\label{eq:rlso}\\
Z_1 &=& 1 + (1 - a^2/M^2)^{1/3} \times
\nonumber\\
& & \left[(1 + a/M)^{1/3} + (1 - a/M)^{1/3}\right]\;,
\\
Z_2 &=& \sqrt{Z_1^2 + 3a^2/M^2}\;.
\end{eqnarray}

Our procedure for building an inspiral and plunge trajectory reduces,
in the equatorial circular limit, to the following algorithm:

\begin{enumerate}

\item Choose a mass $\mu$ for the smaller body, and mass $M$ and spin
$a$ for the larger black hole\footnote{As is common with codes of this
sort, we normalize most dimensionful quantities to $M$.  As such, we
really pick mass ratio $\mu/M$; the impact of $M$ can then be
accounted for in post-processing after the numerics have been
evaluated.}.  Pick an initial orbital radius $r$ and an initial
azimuth $\phi$ for the smaller body's trajectory.

\item Evolve through a sequence of circular, equatorial orbits using
$\dot E$ computed with the frequency-domain code (described in the
following section).  For these orbits, we don't need to compute $\dot
Q$ since $Q = 0$ over the entire sequence.  Also, in this case $\dot
L_z$ is simply related to $\dot E$, so computing it does not provide
additional information.

\item As we approach the last stable orbit, switch to the transition
trajectory following Ref.\ {\cite{s2008}}.  In particular, Ref.\
{\cite{s2008}} describes how to choose the times at which we start and
end the transition regime, which depends in detail on the system's
mass ratio, the larger black hole's spin, and the orbit geometry.  For
circular, equatorial orbits, this prescription reduces to that given
in Ref.\ {\cite{ot2000}}.

\item When we reach $t_{\rm end}$, hold the parameters $E$ and $L_z$
constant, and allow the small body to follow the plunging trajectory
so defined into the larger black hole.

\end{enumerate}

An example trajectory is shown in Fig.\ {\ref{fig:traj}}.  For this
figure, we examine a binary with a mass ratio $\mu/M = 10^{-4}$.  The
larger black hole has a spin $a = 0.3M$.  We start our prograde
trajectory at $r = 5.23M$, close to the LSO at $r = 4.98M$.  The
smaller body orbits roughly 25 times before crossing the LSO; shortly
thereafter, it rapidly plunges into the black hole, locking onto the
horizon as seen by distant observers.  The inset in the left panel of
Fig.\ {\ref{fig:traj}} zooms in on its approach to the horizon,
showing that our plunge trajectory smoothly asymptotes to the final
``horizon locking'' behavior.

\section{Computing radiation}
\label{sec:radiation}

We compute gravitational radiation from our model binaries using the
Teukolsky equation, which describes the evolution of curvature
perturbations to a Kerr black hole \cite{t73,TPRL}. In Boyer-Lindquist
coordinates, it is given by
\begin{eqnarray}
\label{teuk0}
&&
-\left[\frac{(r^2 + a^2)^2 }{\Delta}-a^2\sin^2\theta\right]
         \partial_{tt}\Psi
-\frac{4 M a r}{\Delta}
         \partial_{t\phi}\Psi \nonumber \\
&&- 2s\left[r-\frac{M(r^2-a^2)}{\Delta}+ia\cos\theta\right]
         \partial_t\Psi\nonumber\\  
&&
+\,\Delta^{-s}\partial_r\left(\Delta^{s+1}\partial_r\Psi\right)
+\frac{1}{\sin\theta}\partial_\theta
\left(\sin\theta\partial_\theta\Psi\right)+\nonumber\\
&& \left[\frac{1}{\sin^2\theta}-\frac{a^2}{\Delta}\right] 
         \partial_{\phi\phi}\Psi
+\, 2s \left[\frac{a (r-M)}{\Delta} + \frac{i \cos\theta}{\sin^2\theta}
\right] \partial_\phi\Psi\nonumber\\
&&- \left(s^2 \cot^2\theta - s \right) \Psi = -4\pi\left(r^2+a^2\cos^2\theta\right)T  ,
\label{eq:teuk}
\end{eqnarray}
where $M$ is the mass of the black hole, $a$ its angular momentum per
unit mass, $\Delta = r^2 - 2 M r + a^2$, $r_\pm = M \pm \sqrt{M^2 -
a^2}$.  The quantity $s$ is the ``spin weight'' of the field under
study.  Choosing $s = 0$ means the perturbing field is a scalar field;
$s = \pm 1$ describes a spin-1 (electromagnetic) perturbation, and $s
= \pm 2$ describes gravitational perturbations.  For $s = +2$, the
field $\Psi$ is given by the Weyl curvature scalar $\psi_0$ (see
{\cite{chandra}} for precise definitions and discussion of this
quantity); for $s = -2$, $\Psi = (r - ia\cos\theta)^4\psi_4$, where
$\psi_4$ is another curvature scalar.  We use $s = -2$ since $\psi_4$
is a natural choice to study outgoing radiation.  Once $\psi_4$ is
known, we then know the GWs that the binary produces, since
\begin{equation}
\psi_4 \to \frac{1}{2}\left(\frac{\partial^2 h_+}{\partial t^2} - i
\frac{\partial^2 h_\times}{\partial t^2}\right)
\label{eq:psi4hphm}
\end{equation}
far from the black hole.

The $T$ on the right hand side of Eq.\ (\ref{eq:teuk}) is a source
term constructed from the stress-energy tensor describing a point-like
body moving in the Kerr spacetime.  This stress-energy tensor is given
by
\begin{eqnarray}
T_{\alpha\beta} &=& \mu\int u_\alpha u_\beta\,\delta^{(4)}\left[{\bf x} -
{\bf z}(\tau)\right]d\tau
\\
&=& \mu \frac{u_\alpha u_\beta}{\Sigma \dot t\sin\theta}
\delta\left[r - r(t)\right]
\delta\left[\theta - \theta(t)\right]
\delta\left[\phi - \phi(t)\right]\;.
\nonumber
\label{eq:source_stress}\\
\end{eqnarray}
On the top line, ${\bf x}$ denotes an arbitrary spacetime event, ${\bf
z}(\tau)$ describes the worldline that the point-body follows, and
$\tau$ is proper time along that worldline.  On the second line, we
have performed the integral and written the result in terms of the
body's motion in the Boyer-Lindquist coordinates $r$, $\theta$, and
$\phi$, parameterized by coordinate time $t$.  On both lines,
$u_\alpha = dz_\alpha/d\tau$ is the 4-velocity of the body as it moves
along its worldline.

Note in particular the $\dot t \equiv dt/d\tau$ that appears in the
denominator of Eq.\ (\ref{eq:source_stress}).  This is the time-like
component of the smaller body's geodesic motion, described by Eq.\
(\ref{eq:geod4}).  As the small body approaches the horizon, $\dot t
\to \infty$ --- the passage of coordinate time (time as measured by
distant observers) diverges per unit proper time as measured by that
body.  This is the mechanism by which the source term ``redshifts
away'' as the small body falls into the large black hole, smoothly
converting the Teukolsky equation into its homogeneous form.

The source $T$ is constructed from this $T_{\alpha\beta}$ by
projecting onto a tetrad that describes radiation, and then applying a
particular integro-differential operator; see Refs.\ {\cite{t73,
dh2006, skh07}} for detailed discussion of its nature.  For our
purposes here, the key thing to note is that we must construct the
worldline which the small body follows in order to compute the
radiation associated with its motion.  Different approximations are
appropriate to different regimes of the coalescence, which is why we
have developed two rather different codes for solving Eq.\
(\ref{eq:teuk}).  We now briefly summarize the techniques behind these
two codes, and how we use our solutions.

\subsection{Radiation in the frequency domain}
\label{sec:freqdomain}

As was originally found by Teukolsky {\cite{t73}}, Eq.\
(\ref{eq:teuk}) separates.  For $s = -2$, we put
\begin{equation}
\psi_4 = \frac{1}{(r - ia\cos\theta)^4}\int d\omega \sum_{lm}
R_{lm\omega}(r)S_{lm}(\theta)e^{i(m\phi-\omega t)}\;.
\label{eq:psi4_decomp}
\end{equation}
The function $S_{lm}(\theta)$ is a spin-weighted spheroidal harmonic,
and can be constructed by expanding on a basis of spin-weighted
spherical harmonics {\cite{h2000}}.  The function $R_{lm\omega}(r)$ is
found by solving a second-order ordinary differential equation.  Its
limiting behavior is
\begin{equation}
R_{lm\omega}(r\to\infty) \propto Z^\infty e^{i\omega r^*}\;,
\end{equation}
corresponding to purely outgoing radiation far away, and
\begin{equation}
R_{lm\omega}(r\to r_+) \propto Z^H e^{-i k r^*}\;,
\end{equation}
corresponding to purely ingoing radiation on the event horizon.  The
wavenumber $k = \omega - m\omega_+$, where $\omega_+ = a/2Mr_+$ is the
angular velocity of the hole's event horizon.  In both of these
equations, $r^*$ is the so-called ``tortoise coordinate,''
\begin{equation}
r^* = r + \frac{2Mr_+}{r_+ - r_-}\ln\left(\frac{r - r_+}{2M}\right) -
\frac{2Mr_-}{r_+ - r_-}\ln\left(\frac{r - r_-}{2M}\right)\;.
\end{equation}
The ingoing and outgoing solution is thus characterized by the
coefficients $Z^\infty$ and $Z^H$.  For details of how we compute
these numbers, see Refs.\ {\cite{h2000,dh2006}}.

This frequency-domain approach to solving Eq.\ (\ref{eq:teuk}) is most
useful when the source $T$ has a discrete frequency spectrum.  The
function $\psi_4$ can then be written as a sum over harmonics of the
source's fundamental frequencies.  This is the case for geodesic black
hole orbits; see Ref.\ {\cite{dh2006}} for extensive discussion.

For the circular, equatorial case, orbits and hence the source $T$ are
completely characterized by the frequency $\Omega_\phi$ defined in
Eq.\ (\ref{eq:omega_phi}).  The frequency $\omega$ in Eq.\
(\ref{eq:psi4_decomp}) becomes $m\Omega_\phi$, and the coefficients
$Z^\infty$ and $Z^H$ are then determined by $\omega$ and the harmonic
indices $l$ and $m$.  Once those coefficients are known, it is not
difficult to compute the rates of change of $E$, $L_z$, and $Q$.  The
coefficients $Z^{\infty,H}$ are labeled by the indices $l$ and $m$,
and we have
\begin{eqnarray}
\dot E^{\infty} &=&
\sum_{lm}\frac{|Z^{\infty}_{lm}|^2}{4\pi\omega_m^2}\;,
\label{eq:edot_freq}\\
\dot L_z^{\infty} &=&
\sum_{lm}\frac{m |Z^{\infty}_{lm}|^2}{4\pi\omega_m^3}\;,
\label{eq:lzdot_freq}
\end{eqnarray}
where $\omega_m = m\Omega_\phi$.  Strictly speaking, the $l$ sum
appearing here is from $l = 2$ to infinity, and $m$ is from $-l$ to
$l$; in practice, the sums converge to double precision accuracy once
$l$ is of order a few to a few dozen, depending on how fast the
smaller body orbits.  See Refs.\ {\cite{h2000,dh2006}} for extensive
discussion of convergence issues, as well as for discussion of how to
compute the down-horizon contribution to the rates of change.  Also,
see Ref.\ {\cite{sagoetal}} for discussion of how to compute the rate
of change of $Q$.

\subsection{Radiation in the time domain}
\label{sec:timedomain}

Because they work best when the source has a discrete frequency
spectrum, we only use frequency-domain techniques to describe the
inspiral, when the system is accurately described as slowly evolving
through a sequence of orbits.  When this description is not accurate
(such as in the final plunge, or when inspiral is sufficiently rapid
that the system does not spend many cycles near a given orbit), Eq.\
(\ref{eq:psi4_decomp}) is ill-suited to describing solutions of the
Teukolsky equation.  To handle this case, we solve Eq.\
(\ref{eq:teuk}) directly in the time domain.

In the code we have developed for this, we take advantage of the Kerr
spacetime's axial symmetry to write the field $\Psi$ as {\cite{skh07,
skhd08}}
\begin{eqnarray}
\Psi(t,r,\theta,\phi) & = & \sum_m e^{im\phi}r^3\Phi_m(t,r,\theta)\;.
\label{eq:Phi_def}
\end{eqnarray}
Equation (\ref{eq:teuk}) is then solved as a (2+1)-dimensional partial
differential equation for the modes $\Phi_m$.

The major difficulty in numerically solving Eq.\ (\ref{eq:teuk}) is
coming up with a good description of the source term.  One challenge
is to represent a point-like source on a numerical grid [we use finite
difference techniques to solve Eq.\ (\ref{eq:teuk})].  In Refs.\
{\cite{skh07}} and {\cite{skhd08}}, we have developed a discrete
representation of a delta function which works very well on a
finite-difference grid.  This function is defined so that our
representation of the delta function and of its first two derivatives
preserves various integral identities.  For cases in which a
comparison can be made (e.g., for non-evolving generic geodesic
orbits), we find that this representation allows us to compute GWs in
the time domain with less than a $1\%$ error compared to a
frequency-domain code over a large span of orbital parameter space.

We use the transition and plunge trajectory described in the previous
section to provide the worldline ${\bf z}(\tau)$ and 4-velocity
$u_\alpha$.  As we have already highlighted, the Teukolsky equation
(\ref{eq:teuk}) becomes homogeneous at late times thanks to the manner
in which $\dot t \to \infty$ as the infalling body approaches the
event horizon.  When $T = 0$, the solutions of Eq.\ (\ref{eq:teuk})
are the larger black hole's quasi-normal modes modes.  This means that
the late-time solution in our coalescence model is dominated by
quasi-normal modes of the larger black hole.  By virtue of arising in
a natural way from the behavior of our source term, these modes are
properly phase connected to the preceding inspiral and plunge waves.

\section{Computing recoil from radiation}
\label{sec:recoil}

Once we have computed $\psi_4$, it is not difficult to compute the
rate at which linear momentum is carried by the waves.  Letting
$T_{\alpha\beta}^{\rm GW}$ denote the Isaacson {\cite{isaacson}}
stress-energy tensor for GWs, we have
\begin{eqnarray}
\frac{dP^i(t)}{dt} &=& \lim_{r\rightarrow\infty}r^2\int
n^i\,T^{\rm GW}_{tt}\,d\Omega
\nonumber\\
&=& \lim_{r\rightarrow\infty}\frac{r^2}{16\pi}\int n^i
\left[\left(\frac{\partial h_+}{\partial t}\right)^2 +
\left(\frac{\partial h_\times}{\partial t}\right)^2\right]\,d\Omega
\nonumber\\
&=& \lim_{r\rightarrow\infty}\frac{r^2}{16\pi}\int n^i
\left[\left(\frac{\partial h_+}{\partial t} - i
\frac{\partial h_\times}{\partial t}\right)\cdot\right.
\nonumber\\
& &\left.\qquad\qquad\qquad\qquad
\left(\frac{\partial h_+}{\partial t} + i
\frac{\partial h_\times}{\partial t}\right)\right]\,d\Omega
\nonumber\\
&=& \lim_{r\rightarrow\infty}\frac{r^2}{4\pi}\int n^i
\left|\int_{-\infty}^t \psi_4\,dt'\right|^2\,d\Omega\;.
\label{eq:GWPdot}
\end{eqnarray}
The quantity $n^i$ denotes the Cartesian direction vector in the
large radius limit,
\begin{eqnarray}
n^x &\to& \sin\theta\cos\phi\;,
\\
n^y &\to& \sin\theta\sin\phi\;,
\\
n^z &\to& \cos\theta\;.
\end{eqnarray}
This quantity must then be integrated over time to find the momentum
carried by the GWs:
\begin{equation}
P^i(t) = \int_{-\infty}^t \frac{dP^i(t')}{dt}dt'\;.
\label{eq:GWP}
\end{equation}
Imposing global conservation of momentum, the recoil velocity of the
system is then given by
\begin{equation}
v_{\rm rec}^i(t) = -P^i(t)/M\;.
\label{eq:vrec}
\end{equation}
Equation (\ref{eq:vrec}) will be our primary tool for computing black
hole kicks in this analysis.

For the inspiral, which we model using frequency-domain methods, these
formulas reduce to fairly simple results.  In the $r \to \infty$
limit,
\begin{equation}
\psi_4 = \frac{1}{r}\sum_{lm} Z^\infty_{lm} S_{lm}(\theta)
e^{im(\phi - \Omega_\phi t)}\;.
\end{equation}
Inserting this expansion, the momentum flux formula (\ref{eq:GWPdot})
reduces to a sum over overlap integrals between different modes of the
radiation field.  This integral sharply constrains the mode numbers
which contribute to this formulas sum.  For Schwarzschild, the
integral over $\theta$ can be expressed as a Clebsch-Gordan
coefficient, and we find $l' \in \left[l - 1, l, l + 1\right]$; a
similar but more complicated result describes the integral for Kerr.
For any spin, we find $m' = m \pm 1$ for $P^x(t)$ and $P^y(t)$
[$P^z(t) = 0$ for the equatorial orbits we consider here].  Details of
this calculation will be presented in a separate analysis
{\cite{FDrecoil}}.  For the final plunge and merger portions of the
coalescence, we simply evaluate Eqs.\ (\ref{eq:GWPdot}),
(\ref{eq:GWP}), and (\ref{eq:vrec}) using the $\psi_4$ computed with
our time-domain code.

\section{Results: The coalescence waveform and recoil}
\label{sec:results}

We now put the pieces of this formalism together to compute the
waveforms from binary black hole coalescence and to calculate recoil.
We begin by describing some issues with extrapolating from the truly
perturbative mass ratios we study here (Sec.\ {\ref{sec:massratio}}),
and then describe in more detail how we assemble the full inspiral
trajectory and its associated waveform (Sec.\ {\ref{sec:waves}})
before discussing our results for the recoil (Sec.\
{\ref{sec:recoil}}).  As already mentioned, we focus in this analysis
on quasi-circular equatorial configurations.  We conclude (Sec.\
{\ref{sec:converge}}) by describing the convergence of our recoil
results.  We find that as we go to large spin, prograde mergers may
require a large number of $m$ modes [cf.\ the axial decomposition
(\ref{eq:Phi_def}) we use] to give convergent results.

\subsection{Mass ratio dependence considerations}
\label{sec:massratio}

By using the Teukolsky equation to model coalescence, we are by
construction working to first order in mass ratio --- the curvature
scalar $\psi_4$ that we compute neglects all corrections of order
$(\mu/M)^2$.  Since the various fluxes we compute (energy, momentum,
angular momentum) follow from the modulus squared of $\psi_4$, it
likewise follows that these fluxes are all strictly proportional to
$(\mu/M)^2$.  The recoil velocity, as an integral of the momentum
flux, should likewise scale essentially with $(\mu/M)^2$.  We may
expect small deviations from this scaling since the timescales of
inspiral and of plunge and merger do not scale with mass ratio in
quite the same way.  However, we have found that a $(\mu/M)^2$ scaling
describes the final recoil very accurately for all mass ratios more
extreme than $\mu/M = 10^{-3}$; we have not examined mass ratios less
extreme than this yet.  Since the scaling with mass ratio is trivial,
we will present detailed results for only one choice, $\mu/M =
10^{-4}$.  In our summary figure for total recoil velocity as a
function of spin (Fig.\ {\ref{fig:kick}}), we normalize our results by
the scaling $(\mu/M)^2$.

In Ref.\ {\cite{FHH}}, it was argued that one can improve the ability
of perturbation theory to extrapolate out of the perturbative regime
by replacing the $(\mu/M)^2$ which describes the momentum flux and the
recoil velocity with
\begin{equation}
f(\mu/M) = \left(\frac{\mu}{M}\right)^2\sqrt{1 - \frac{4\mu}{M}}\;.
\end{equation}
In this argument, it is claimed that in extrapolating out of the
perturbative regime it is useful to interpret the small body's mass
$\mu$ as the system's reduced mass, and the large black hole's mass
$M$ as the system's total mass.  A similar interpretation of these
masses has been shown to give excellent results interpreting the
head-on collisions of black holes {\cite{smarr}}.  The function
$f(\mu/M)$ has a maximum $f_{\rm max} = 0.01789$ at $\mu/M = 0.2$
(corresponding, after remapping the meaning of these mass parameters,
to $m_{\rm small}/m_{\rm large} = 0.382$).

As we will discuss in more detail later in this section, using this
scaling doesn't work quite as well as we might have hoped.  The key
issue is that in our perturbative framework, we assume there exists a
stationary background spacetime which we can expand around, and that
this background does not evolve during the coalescence.  This means,
for example, that a binary which contains a large Schwarzschild black
hole at early times will evolve to a single Schwarzschild black hole
at late times; we fail to account for the evolution of this black
hole's spin during the merger.  This is a minor error when the mass
ratio is small, but is significant for large mass ratio.  In
particular, for mass ratios $\mu/M \sim 0.1$ or larger, the spin of
the final black hole will change substantially in the merger.  By not
evolving the spin properly, we do not get the late time spectrum of
merger/ringdown waves correct, with important consequences for the
system's final kick.

\subsection{Example waveform}
\label{sec:waves}

\begin{figure}[hbt]
\includegraphics[width = 8.6cm]{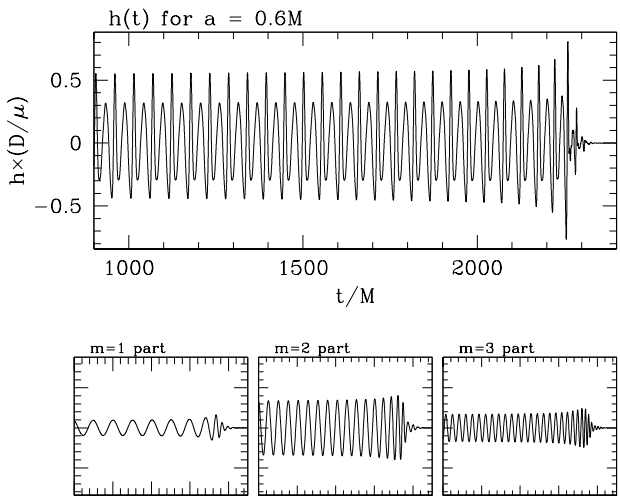}
\vskip -1.5cm
\caption{Coalescence waveform computed with perturbation theory for a
binary with mass ratio $\mu/M = 10^{-4}$, and in which the larger
black hole has spin $a = 0.6M$.  We show the $+$ polarization of the
waveform as viewed in the binary's equatorial plane; the $\times$
waveform is zero from this viewing angle.  The wave is normalized by
$D$, the distance from source to observer, and the origin of the time
axis is arbitrary.  The waveform shown in the top panel includes
contributions from all modes with $|m| \le 6$; the individual
contributions for $m = 1$, $m = 2$, and $m = 3$ are shown below.
Notice how the smoothly chirping ``inspiral'' waves blend naturally
into the rapidly damped ``ringdown'' which terminates this waveform.}
\label{fig:h0.6}
\end{figure}

\begin{figure}[hbt]
\includegraphics[width = 8.6cm]{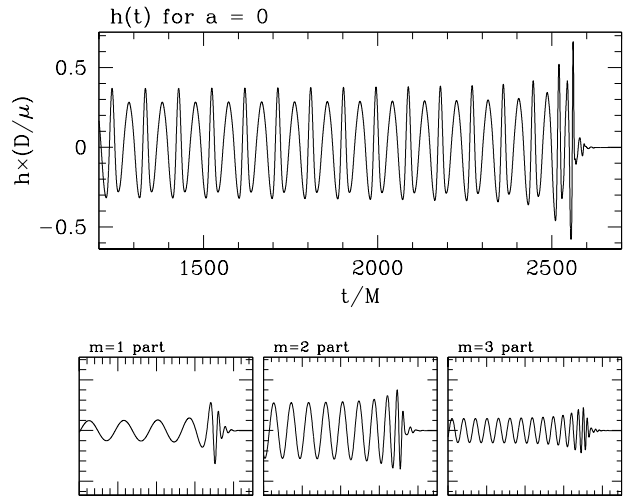}
\vskip -1.5cm
\caption{Same as Fig.\ {\ref{fig:h0.6}}, but the large black hole has
spin $a = 0$ in this case.  The frequencies which describe this wave
are generically lower than those shown in Fig.\ {\ref{fig:h0.6}} since
the transition from inspiral to plunge happens at larger radius thanks
to smaller spin parameter in this binary.  In addition, the final
ringdown waves damp out more rapidly than in the $a = 0.6M$ case.}
\label{fig:h0}
\end{figure}

\begin{figure}[hbt]
\includegraphics[width = 8.6cm]{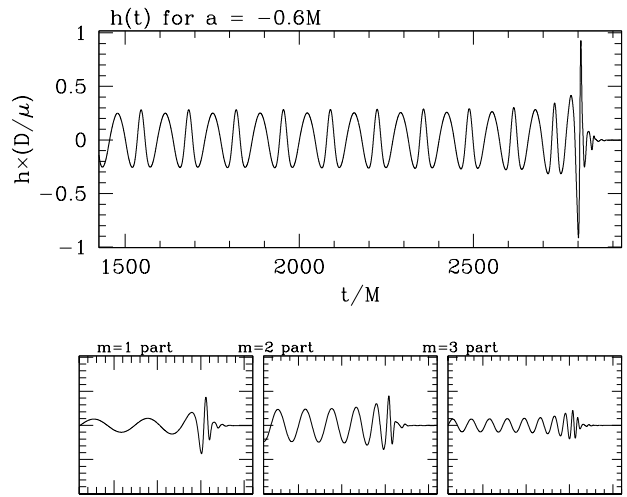}
\vskip -1.5cm
\caption{Same as Figs.\ {\ref{fig:h0.6}} and {\ref{fig:h0}}, but
now for black hole spin $a = -0.6M$ (i.e., same black hole as in Fig.\
{\ref{fig:h0.6}}, but now for a retrograde orbit geometry).  The
frequencies characterizing this wave are again lower than for the two
previous examples, and the ringdown waves damp even more rapidly.}
\label{fig:hm0.6}
\end{figure}

Figures {\ref{fig:h0.6}}, {\ref{fig:h0}}, and {\ref{fig:hm0.6}}
present coalescence waveforms for a binary with $\mu/M = 10^{-4}$, and
in which the larger black hole has spin $a/M = 0.6$, $0$, and $-0.6$
respectively.  We focus on the late waves, including the final plunge
and ringdown.  The data for these figures were generated using the
time-domain code discussed in Sec.\ {\ref{sec:timedomain}}.  In the
largest panel, we show the wave including all contributions with $|m|
\le 6$; the three smaller panels show individual contributions from
the $m = 1$, $2$, and $3$ modes.

In all three examples we show, the general character of the waveforms
is essentially the same: a slowly evolving chirping sinusoid that
terminates in an exponentially damped ringdown.  Two aspects of the
waveforms clearly differ as we move from $a = 0.6M$ to $a = -0.6M$.
First, notice that the waveform for $a = 0.6M$ is clearly of rather
higher frequency than for $a = 0$, which in turn is higher than for $a
= -0.6M$.  This is not surprising, and is a simple consequence of the
orbit's geometry: the LSO, which approximately delineates the
transition from inspiral to final plunge, is at $r_{\rm LSO} = 3.83M$
for $a = 0.6M$, $r_{\rm LSO} = 6M$ for $a = 0$, and $r_{\rm LSO} =
7.85M$ for $a = -0.6M$.  As the LSO moves to larger radius, the
orbital frequency associated with it sweeps lower.

Second, the final ringdown waves damp more quickly as we move from the
prograde to the retrograde configuration.  This is also not
surprising, and follows naturally from the damping behavior of a Kerr
black hole's quasi-normal modes: modes which are ``parallel'' to a
hole's spin (i.e., have $m > 0$ for $a > 0$, and vice versa) are much
more long lived than ``antiparallel'' modes.  See, for example, Fig.\
45 of Ref.\ {\cite{chandra}} (noting that Chandra's sign convention on
the Fourier transform means that $m_{\rm Chandra} = -m_{\rm us}$).

\subsection{Recoil versus spin}
\label{sec:spin}

\begin{figure}[hbt]
\includegraphics[width = 8.6cm]{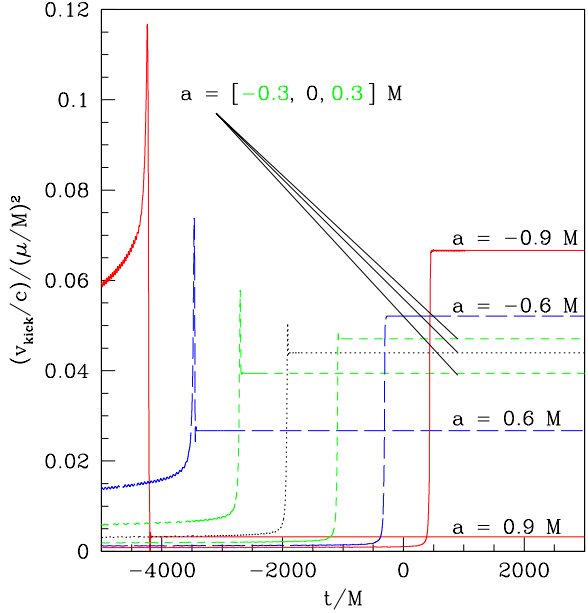}
\caption{Summary of recoil versus time for large black hole spins $a/M
\in [-0.9, -0.6, -0.3, 0, 0.3, 0.6, 0.9]$.  For each track, the time
origin is arbitrary, so we have shifted the data in order to cleanly
display all seven recoil trends shown here.  The key feature we find
is the manner in which (especially for large spin, prograde
coalescences) the kick builds to a large positive value, followed by
an ``antikick'' that brings the total accumulated recoil down to much
smaller values.  The antikick is especially strong when the spin is
large and the coalescence is prograde, and is essentially non-existent
for large spin retrograde coalescence.}
\label{fig:kick}
\end{figure}

Figure {\ref{fig:kick}} summarizes how the kick imparted to a binary
behaves as a function of spin for mass ratio $\mu/M = 10^{-4}$.  In
this plot, we show the magnitude of the recoil that has accumulated up
to some time $t$.  Since the origin of the time axis is not
particularly interesting, we have shifted the various tracks so that
we can easily compare how the recoil varies as a function of spin.

The clearest feature apparent here is that, especially for prograde
coalescences ($a > 0$), the recoil grows to some large positive value,
but then is strongly suppressed by an ``antikick'' to something
significantly smaller.  The suppression is a very strong function of
the large black hole's spin: For the five cases which show antikick
behavior in Fig.\ {\ref{fig:kick}}, the peak kick $v^{\rm peak}$ and
the late time kick $v^{\rm late}$ are given by
\begin{eqnarray}
\mbox{$a = 0.9M$:}\qquad
v^{\rm peak}/c &=& 0.17\,(\mu/M)^2
\nonumber\\
v^{\rm late}/c &=& 0.0032\,(\mu/M)^2\;;
\end{eqnarray}
\begin{eqnarray}
\mbox{$a = 0.6M$:}\qquad
v^{\rm peak}/c &=& 0.074\,(\mu/M)^2
\nonumber\\
v^{\rm late}/c &=& 0.027\,(\mu/M)^2\;;
\end{eqnarray}
\begin{eqnarray}
\mbox{$a = 0.3M$:}\qquad
v^{\rm peak}/c &=& 0.058\,(\mu/M)^2
\nonumber\\
v^{\rm late}/c &=& 0.039\,(\mu/M)^2\;;
\end{eqnarray}
\begin{eqnarray}
\mbox{$a = 0$:}\qquad
v^{\rm peak}/c &=& 0.051\,(\mu/M)^2
\nonumber\\
v^{\rm late}/c &=& 0.044\,(\mu/M)^2\;;
\end{eqnarray}\begin{eqnarray}
\mbox{$a = -0.3M$:}\qquad
v^{\rm peak}/c &=& 0.048\,(\mu/M)^2
\nonumber\\
v^{\rm late}/c &=& 0.047\,(\mu/M)^2\;.
\end{eqnarray}
In other words, we find that the antikick suppresses the maximum
recoil by a factor of $53$ for $a = 0.9M$, $2.7$ for $a = 0.6M$, $1.5$
for $a = 0.3M$, $1.2$ for $a = 0$, and $1.02$ for $a = -0.3M$.  The
late-time kick shown in Fig.\ {\ref{fig:kick}} is nicely fit by the
formula
\begin{eqnarray}
v_{\rm rec}(a)/c &\simeq& \left[0.0440 - 0.0099(a/M) - 0.0114 (a/M)^2
\right.
\nonumber\\
& & \left. - 0.0312 (a/M)^3\right](\mu/M)^2\;.
\end{eqnarray}
Over most of the relevant parameter space, this comes in right between
the ``upper'' and ``lower'' estimates of Ref.\ {\cite{FHH}} [compare
to Eqs.\ (1) and (2) of Ref.\ {\cite{MMFHH}}].

The antikick behavior we see agrees at least qualitatively with the
trends seen in Schnittman et al.\ {\cite{antikick}}.  It is hard to
calibrate the quantitative agreement between these analyses, since (as
discussed above in Sec.\ {\ref{sec:massratio}}) extrapolating from the
perturbative regime into that of the mass ratios considered in Ref.\
{\cite{antikick}} is not as simple as the arguments in Ref.\
{\cite{FHH}} would suggest.  Consider the Schwarzschild coalescence
results.  If we use the $(\mu/M)^2 \to f(\mu/M)$ rule suggested in
Ref.\ {\cite{FHH}}, we find
\begin{eqnarray}
v^{\rm late}(a = 0) &=& 0.044 f(\mu/M)c
\nonumber\\
&\le& 235\,{\rm km/sec}\;.
\end{eqnarray}
On the second line, we have used $f_{\rm max} = 0.01789$ in order to
estimate how large the kick can be in this case.

Taken at face value, this suggests that the recoil of Schwarzschild
black holes has a maximum of 235 km/sec, $34\%$ higher than the
maximum value of 175 km/sec found in careful numerical relativity
calculations {\cite{gsbhh07}}.  However, in those numerical relativity
calculations, the final black hole is not Schwarzschild, but has a
spin $a \simeq 0.67M$.  Figure {\ref{fig:kick}} tells us we should
expect a larger ``antikick'' when the final black hole is rapidly
spinning.  Applying the same extrapolation to our recoil data for $a =
0.6M$ (the nearest value to $a = 0.67M$ in our dataset) leads to
\begin{eqnarray}
v^{\rm late}(a = 0.6M) &=& 0.027 f(\mu/M)c
\nonumber\\
&\le& 144\,{\rm km/sec}\;.
\end{eqnarray}
This is about $18\%$ lower than the numerical relativity prediction.
The lesson we take from this is that naive extrapolation from the
small mass ratio regime does not give a good estimate of the final
kick in the comparable mass case.  Because the background spacetime is
fixed, we do not accurately describe the system's final state and
hence the last waves that it emits during the coalescence.

\subsection{Convergence}
\label{sec:converge}

As discussed in Sec.\ {\ref{sec:timedomain}}, our time-domain
perturbation theory code expands the field $\Psi$ in axial modes; cf.\
Eq.\ (\ref{eq:Phi_def}).  The angular integral in Eq.\
(\ref{eq:GWPdot}) takes the form
\begin{eqnarray}
\frac{dP_x}{dt} &\propto& \sum_{m,m'} \int_0^{2\pi}
d\phi\,\cos\phi e^{i(m - m')\phi} \Phi_m\Phi^*_{m'}
\nonumber\\
&\propto& \sum_{m,m'} \left(\delta_{(m+1),m'} +
\delta_{(m-1),m'}\right)\Phi_m\Phi^*_{m'}\;,
\label{eq:Pdotx_mode}\\
\frac{dP_y}{dt} &\propto& \sum_{m,m'} \int_0^{2\pi}
d\phi\,\sin\phi e^{i(m - m')\phi} \Phi_m\Phi^*_{m'}
\nonumber\\
&\propto& \sum_{m,m'} \left(\delta_{(m+1),m'} -
\delta_{(m-1),m'}\right)\Phi_m\Phi^*_{m'}\;.
\label{eq:Pdoty_mode}
\end{eqnarray}
Contributions from terms with $m = m'$ vanish; the recoil arises from
beating between adjacent $m$-modes.

The question we now address is how many $m$-modes must be included in
order to accurately compute the recoil.  We have found that this is a
strong function of black hole spin: When the black hole has large
positive spin, many more modes are needed for convergence than for
small or retrograde coalescence.

Tables {\ref{tab:a=m0.6_conv}} -- {\ref{tab:a=0.6_conv}} summarize
convergence data for three of the cases presented in Fig.\
{\ref{fig:kick}}.  We show, as a function $m_{\rm max}$ [the value of
$m$ and $m'$ at which the sums in Eqs.\ (\ref{eq:Pdotx_mode}) and
(\ref{eq:Pdoty_mode}) are terminated] the peak magnitude of the
momentum flux normalized by $(\mu/M)^2$,
\begin{equation}
\dot {\cal P} \equiv \left[\frac{\sqrt{(dP_x/dt)^2 +
(dP_y/dt)^2}}{(\mu/M)^2}\right]_{\rm max}\;.
\end{equation}
The ``max'' subscript means that we select the maximum of this
quantity over the timespan for which we compute the momentum flux.
This quantity is given in units of $M^{-1}$.  We also show the
percentage change in $\dot{\cal P}$ as we increase $m_{\rm max}$ by
one.

\begin{table}[ht]
\caption{Convergence of recoil with azimuthal mode for $a = -0.6M$.
First column is $m_{\rm max}$, the largest value of $m$ we include.
Second column is the value of $\dot{\cal P}$, the peak magnitude of
the momentum flux normalized by $(\mu/M)^2$.  The third column gives
the percentage change in $\dot{\cal P}$ we find as we increase $m_{\rm
max}$ from the previous value.}
\begin{tabular}{|c|c|c|}
\hline  
$m_{\rm max}$ & $\dot{\cal P}$ & $\%$ change \\
\hline  
$2$ & $2.855 \times 10^{-3}$ & --- \\
$3$ & $4.030 \times 10^{-3}$ & $29.2\%$ \\
$4$ & $4.557 \times 10^{-3}$ & $11.6\%$ \\
$5$ & $4.807 \times 10^{-3}$ & $5.2\%$ \\
$6$ & $4.930 \times 10^{-3}$ & $2.5\%$ \\
\hline
\end{tabular}
\label{tab:a=m0.6_conv}
\end{table}

\begin{table}[ht]
\caption{Convergence of momentum flux with $m$ for $a = 0$.  All
details are as in Tab.\ {\ref{tab:a=m0.6_conv}}.}
\begin{tabular}{|c|c|c|}
\hline  
  $m_{\rm max}$ & $\dot{\cal P}$ & $\%$ change \\    
\hline  
$2$ & $1.712 \times 10^{-3}$ & --- \\
$3$ & $4.188 \times 10^{-3}$ & 58.9\%\\
$4$ & $5.508 \times 10^{-3}$ & 24.0\%\\
$5$ & $6.182 \times 10^{-3}$ & 10.9\%\\
$6$ & $6.532 \times 10^{-3}$ & 5.4\%\\
\hline
\end{tabular}
\label{tab:a=0_conv}
\end{table}

\begin{table}[ht]
\caption{Convergence of momentum flux with $m$ for $a = 0.6M$.  All
details are as in Tab.\ {\ref{tab:a=m0.6_conv}}.}
\begin{tabular}{|c|c|c|}
\hline  
  $m_{\rm max}$ & $\dot{\cal P}$ & $\%$ change \\    
\hline  
$2$ & $1.373 \times 10^{-3}$ & --- \\
$3$ & $7.488 \times 10^{-3}$ & 81.7\%\\
$4$ & $1.105 \times 10^{-2}$ & 32.2\%\\
$5$ & $1.302 \times 10^{-2}$ & 15.1\%\\
$6$ & $1.412 \times 10^{-2}$ & 7.8\%\\
\hline
\end{tabular}
\label{tab:a=0.6_conv}
\end{table}

Tables {\ref{tab:a=m0.6_conv}} -- {\ref{tab:a=0.6_conv}} indicate
that, once several modes have been computed, the fractional error in
the momentum flux decreases by roughly a factor of two with each unit
increase in $m$.  However, the magnitude of the relative error is a
rather strong function of black hole spin.  For $a = -0.6M$, we find
that going from $m_{\rm max} = 5$ to $m_{\rm max} = 6$ changes the
momentum flux by only $2.5\%$.  Including additional modes presumably
will only produce percent-level changes.  For $a = 0.6M$ by contrast,
the flux changes by nearly $8\%$ as $m_{\rm max}$ is increased from 5
to 6.  Many modes are clearly needed to accurately compute the waves
(and the recoil from these waves) as the large black hole's spin
approaches the Kerr maximum.

\section{Conclusions and future work}
\label{sec:conclude}

Now that numerical relativity has effectively solved the two-body
problem in general relativity, a major task for researchers has become
to explore the parameter space of binary coalescence.  This will
insure that wave models constructed as templates for GW data analysis
fully encompass the range of behaviors that are likely in real binary
mergers, and allow us to more fully understand the phenomenology of
binary black hole merger astrophysics.  In this analysis, we have
demonstrated that perturbation theoretical techniques based on the
Teukolsky equation are an excellent tool for extending the reach of
our computations, allowing us to model large mass ratios that are
challenging for 3+1 numerical simulations, but may be of astrophysical
significance.  Our analysis joins previous work by Damour and
colleagues {\cite{ndt07,dn07}}, Mino and Brink {\cite{mb08}}, and by
Lousto and colleagues {\cite{lnzc}} which likewise used perturbation
theory to model large mass ratio binaries.  By using the Teukolsky
equation, we can explore how the larger black hole's spin impacts the
analysis, exemplified by our demonstration of how the previously
identified ``antikick'' {\cite{antikick}} strongly depends on this
spin.

Two directions for future analysis strike us as particularly
noteworthy.  First, the major motivation for this work is that
perturbation theory makes exploring parameter space computationally
fast and simple.  As such, it would be worthwhile to continue this
exploration, examining how the waveform varies as a function of
spin-orbit alignment, and exploring (for example) how the antikick
evolves as one varies the inclination smoothly from the prograde to
the retrograde geometry.  Preliminary calculations of this behavior
indicate that the antikick rapidly evolves with spin-orbit alignment,
consistent with the results of Mino and Brink {\cite{mb08}} which
demonstrate a strong dependence on the final kick with the plunge
geometry.

Second, as Damour and Nagar have emphasized {\cite{dn07}},
particularly useful application comes by including input from the
effective one-body formalism in our description of the small body's
motion; input from perturbation theory can likewise be used to
calibrate certain parameters in the EOB framework.  Now that the
spin-augmented Hamiltonian for binary systems is understood
{\cite{djs08,brb09}}, we expect that work to extend EOB to more
broadly include the impact of spin will become very active.  We hope
that the tools we have presented here will be useful for further
refining what has already proved to be a valuable tool for modeling
coalescing binaries.

\section*{Acknowledgments}

We gratefully acknowledge helpful correspondence and discussion with
Thibault Damour and Alessandro Nagar regarding the literature on black
hole perturbation theory and the effective one-body approach; we also
thank Thibault Damour for useful feedback on a draft of this paper.
We likewise thank Carlos Lousto and Hiroyuki Nakano for helpful
comments on an earlier version.  A very early draft of this manuscript
was presented as a chapter in the Ph.\ D.\ thesis of PAS
{\cite{pranesh_thesis}}.  This work was supported at MIT by NASA Grant
No.\ NNG05G105G and NSF Grant PHY-0449884.  GK acknowledges research
support from NSF Grants PHY-0831631 and PHY-0902026, and hardware
donations from Sony and IBM.  SAH gratefully acknowledges the support
of the Adam J.\ Burgasser Chair in Astrophysics in completing this
analysis.

\end{document}